\documentclass[%
 reprint,
 amsmath,amssymb,
 aps,
 prl
]{revtex4-2}

\usepackage{graphicx}
\usepackage{dcolumn}
\usepackage{bm}
\usepackage{hyperref}
\usepackage[dvipsnames]{xcolor}

\newcommand{\EndMatter}{\hyperref[endmatter]{End Matter}}

\begin{document}

\preprint{APS/123-QED}

\title{Ultrafast directed transport via energy recuperation in non-Markovian systems}

\author{Mateusz Wi\'{s}niewski}
\author{Jakub Spiechowicz}%
 \email{jakub.spiechowicz@us.edu.pl}
\affiliation{%
 Institute of Physics, University of Silesia, 41-500 Chorz\'{o}w, Poland
}%

\date{\today}

\begin{abstract} 
A recent pioneering experiment [Nat.~Commun.~16, 10114 (2025)] demonstrated that a driven overdamped colloidal particle in a~harmonic trap immersed in a~viscoelastic fluid can recuperate energy dissipated into the surrounding bath and convert it into useful work. In this article we considerably extend the original predictions. In particular, we show that energy recuperation is a generic feature of non-Markovian systems both in and out of equilibrium, even as simple as a free Brownian particle. Moreover, we demonstrate that inertia alone, even in the strong damping regime, can lead to this effect despite the absence of any external forcing. These results suggest that energy recuperation can be ubiquitous in nature and it may be the \emph{modus operandi} of various phenomena in setups with memory. We show that this novel mechanism of energy recovery is the source of memory-induced ultrafast directed transport of a particle in a periodic potential in which it almost attains its top speed corresponding to the system with no energy barriers. Our results may answer from the fundamental point of view the question why the cytosol, the intracellular fluid in biological cells, is viscoelastic.
\end{abstract}

\maketitle

Dynamics of non-Markovian systems is a classic scientific problem, yet it attracts everlasting research activity in physics and beyond. The reason for this is the fact that in reality there are no memoryless setups, but actual systems retain information about their past through temporal correlations \cite{VanKampen1998}. Recent advances in experimental platforms and theoretical methods capable of probing these correlations in complex environments make this issue a hot topic nowadays \cite{Odeh2025, Cao2023, Plissier2026, Engbring2023, Li2025, Dalton2023}.

Memory arises naturally when a system interacts with a bath possessing relaxation times comparable to timescales of its own dynamics. The presence of memory may significantly alter characteristics of the setup, such as the rate of barrier crossing \cite{BarbierChebbah2024, Ginot2022, Ferrer2021}, kinetics of isomerization \cite{Dalton2024} and behavior of active matter \cite{Abbasi2023, Tucci2022}, to name only a few. Memory can also enrich the dynamical response of the system and give rise to phenomena that are absent in corresponding memoryless setups, including absolute negative mobility \cite{Wisniewski2024chaos}, current reversal \cite{Wisniewski2025}, particle recoil \cite{Chapman2014, GomezSolano2015, Pruszczyk2025} and oscillations of overdamped Brownian particles \cite{Berner2018, Venturelli2023}. 

A striking and very timely example of a memory-induced phenomenon is the experimental observation of energy recuperation in non-Markovian baths \cite{Ginot2025}. In viscous fluids the Markovian system irreversibly dissipates heat into its environment. In contrast, viscoelastic baths can store some part of that energy due to their elastic properties \cite{Waigh2005}. The authors of Ref.~\cite{Ginot2025} demonstrated that the non-Markovian system consisting of a driven overdamped colloidal particle can recover a portion of the energy dissipated into the surrounding medium and convert it into useful work. These findings highlight the fundamental role of memory in nonequilibrium energetics. 

In this Letter we establish a microscopic theoretical foundation for this intriguing phenomenon and considerably extend the original predictions. In particular, we show that energy recuperation is a generic feature of non-Markovian systems even as simple as a free Brownian particle, both in and out of equilibrium (see End Matter). Moreover, by investigating a strongly damped system instead of the overdamped one, we demonstrate that inertia alone can lead to this effect despite the absence of any external forcing. Our findings suggest that energy recuperation may be ubiquitous in nature and it might be the \emph{modus operandi} of various phenomena occurring in assistance of memory. This applies also to open quantum systems which are inherently non-Markovian \cite{Spiechowicz2021}.

As an exemplification we show that the discovered mechanism of energy recovery via inertia may induce ultrafast directed transport in non-Markovian periodic systems in the strong damping regime. This result might answer from the fundamental point of view the question why the cytosol, the intracellular fluid in biological cells, is viscoelastic \cite{Bressloff2013}. It turns out that solely non-Markovianity of the intracellular environment can speed up the intracellular transport or even induce it thus optimizing the functioning of the living cell.

We consider a Brownian particle of mass $m$ immersed in a correlated thermal bath and subjected to a potential $U(x)$. We describe its dynamics with a dimensionless Generalized Langevin Equation \cite{Zwanzig1973,Hanggi2007,Wisniewski2024pre1,Wisniewski2024pre2}
\begin{equation} \label{eq:gle}
	m\dot{v}(t) = w(t) - U'(x) + \eta(t),
\end{equation}
where
\begin{equation}
    w(t) = -\gamma\int_0^t K(t-t') v(t') \mathrm{d}t'
\end{equation}
is the retarded friction force. From now on we assume $\gamma = 1$ due to the choice of length and time scales presented in the \EndMatter{}. The bath is characterized by the memory kernel $K(t)$, which we impose as
\begin{equation} \label{eq:K}
	K(t) = 2\varepsilon \delta(t) + \frac{1-\varepsilon}{\tau} e^{-|t|/\tau},
\end{equation}
where $\delta(t)$ is the Dirac delta, $\tau$ is the memory time and $\varepsilon\in[0,1]$ determines the magnitude of the contribution of the fast (unresolved) bath variables to $K(t)$ \cite{Hanggi1978, Zwanzig2001}.
The parameters are chosen such that in the memoryless limit $\tau\to 0$ the kernel reduces to $K_0(t) = 2\delta(t)$ and the bath is purely viscous. 
By virtue of the fluctuation-dissipation theorem \cite{Kubo1966} $K(t)$ defines also the autocorrelation function of thermal noise $\eta(t)$, namely
\begin{equation}
	\langle \eta(t) \eta(t') \rangle = \gamma \theta K(t-t'),
\end{equation}
with $\theta$ standing for the dimensionless temperature of the bath and $\langle \cdot \rangle$ indicating an average over the ensemble of system trajectories.

\begin{figure}
    \centering
    \includegraphics{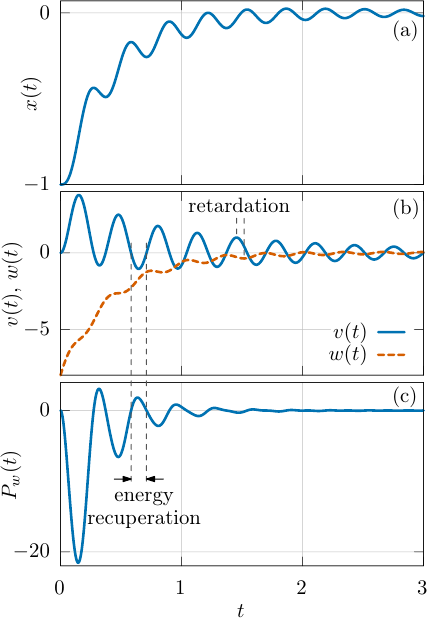}
    \caption{Relaxation of the particle in the harmonic potential $U_h(x)$ starting from $x_0 = -1$ with the initial velocity $v(0) = 0$. Other parameters read $m=0.03$, $\gamma=1$, $\varepsilon=0.01$, $\tau = 0.3$, $k = 8$ and $\theta = 0$. (a): the position $x(t)$, (b): the velocity $v(t)$ and the retarded friction force $w(t)$, (c): the rate of energy exchange between the particle and the bath $P_w(t)$.}
    \label{fig:traj_harmonic}
\end{figure}

\paragraph{Energy recuperation in the relaxation of an inertial non-Markovian system.}
Let us first analyze the process of relaxation of the particle in a harmonic potential $U_h(x) = \frac{1}{2}kx^2$ in the deterministic regime $\theta = 0$. For the memory kernel given by Eq.~(\ref{eq:K}) the Generalized Langevin Equation (\ref{eq:gle}) is equivalent to (for details see \EndMatter{})
\begin{equation} \label{eq:char_eq}
    m\tau \dddot{x} + (\gamma\varepsilon\tau + m)\ddot{x} + (k\tau + \gamma)\dot{x} + kx = 0.
\end{equation}
The corresponding characteristic equation is of third order and its roots $\{\lambda_i,\ i = 1,2,3\}$ can be calculated analytically. Their form is complicated and thus we present them only in two limiting cases.
First, in the memoryless limit $\tau \to 0$ they read
\begin{equation} \label{eq:root_memless}
    \lambda_1^0 = -\frac{1}{\tau},\ \lambda_{2,3}^0 = -\frac{\gamma \pm \sqrt{\gamma^2-4mk}}{2m}.
\end{equation}
In the strong damping regime $\gamma^2 > 4mk$ the particle dynamics is determined by a sum of three exponential decays with the first one corresponding to $\lambda_1^0$ vanishing rapidly. For the initial value problem with $x(0) = x_0$, $v(0) = 0$ and $\dot{v}(0) = 0$, it implies that the particle quickly relaxes towards the potential minimum and no oscillations are observed.
In contrast, in the limit of long memory time $\tau \to \infty$ the roots read
\begin{equation} \label{eq:root_longmem}
	\lambda_1^\infty = -\frac{1}{\tau},\ \lambda_{2,3}^\infty = -\frac{\gamma\varepsilon \pm \sqrt{(\gamma\varepsilon)^2 - 4mk}}{2m}.
\end{equation}
Note that now in the same strong damping regime $\gamma^2 > 4mk$, the first root $\lambda_1^\infty$ is again real, but the other two $\lambda_{2,3}^\infty$ may be complex when $(\gamma\varepsilon)^2 < 4mk$. It implies that the particle relaxation is oscillatory, see Fig.~\ref{fig:traj_harmonic}(a). This means that the memory induces oscillations that would not appear in the memoryless system. Oscillating solutions have already been described in the literature even for overdamped (inertialess) particles \cite{Berner2018, Venturelli2023}, however, it was observed only in driven systems. We show that for inertial setups in principle it can be detected even in the relaxation process with no external forcing.

The presence of oscillations can be intuitively explained by the analysis of the friction force $w(t)$ experienced by the particle. In Fig.~\ref{fig:traj_harmonic}(b) we present an example trajectory $w(t)$ obtained for $m=0.03$, $\gamma=1$, $\varepsilon=0.01$, $k = 8$, $x_0 = -1$ and $\tau = 0.3$, for which the particle in the memoryless limit is strongly damped. 
For purely viscous bath $\tau = 0$ the friction force $-\gamma v(t)$ is always opposite and proportional to the instantaneous particle velocity $v(t)$. The presence of memory $\tau \neq 0$ implies that the medium needs time of order $\tau$ to build up its response, so often the effective friction experienced by the particle is smaller than in the viscous case and therefore the system is less damped.

The emergence of oscillatory motion crucially affects the exchange of energy between the particle and the surrounding bath. In the deterministic limit $\theta=0$ the interaction between the environment and the particle is determined solely by the friction force $w(t)$ and the rate of energy exchange reads $P_w(t) = w(t)v(t)$, with $P_w(t) < 0$ indicating transfer of energy from the particle to the environment, and $P_w(t) > 0$ representing the reversed scenario, i.e.~the recuperation of energy from the bath by the particle. In the memoryless limit $P_w(t) = -\gamma v^2(t)$ and the particle only dissipates energy to the environment. 

In the non-Markovian system, however, the friction force $w(t)$ is retarded with respect to the velocity and when $v(t)$ changes its sign due to the oscillatory motion, the energy exchange rate $P_w(t)$ can become positive. Indeed in Fig.~\ref{fig:traj_harmonic}(b) one can find time periods when both $v(t)$ and $w(t)$ are of the same sign, which corresponds to positive $P_w(t)$ in Fig.~\ref{fig:traj_harmonic}(c). The recuperated energy is converted to the particle potential and kinetic energy, so unlike in the memoryless case, the bath can temporarily perform positive work on the system. Our discussion so far shows that inertia alone can lead to energy recuperation in non-Markovian dynamics, in particular in strong damping regime, even in absence of any external forces.

\paragraph{Ultrafast directed transport induced by energy recuperation.}
The process of particle relaxation in the harmonic potential $U_h(x)$ can be mapped onto an escape problem which is fundamental for directed transport in a periodic potential with energetic barriers \cite{Risken1996}. 
The emergence of energy recuperation in the presence of memory suggests that in non-Markovian system the particle can overcome 
the potential barrier, even though it could not do it in the purely viscous case of the memoryless regime. In the context of the periodic potential it would result in the emergence of running solutions of Eq.~(\ref{eq:gle}) and consequently the occurrence of directed transport.

\begin{figure}
    \centering
    \includegraphics{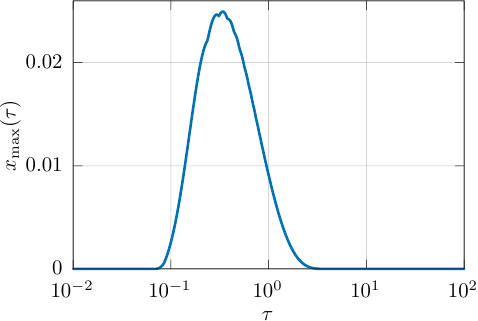}
    \caption{
    The maximal position $x_\mathrm{max}$ of the particle starting from $x_0=-1$ with the initial velocity $v(0) = 0$ in the harmonic potential $U_h(x)$ as a function of the memory time $\tau$. Other parameters are the same as in Fig.~\ref{fig:traj_harmonic}.} 
    \label{fig:v_xmax_tau}
\end{figure}

To test this hypothesis we first investigate the maximal position $x_\mathrm{max}$ of the particle in the harmonic potential $U_h(x)$ for the same parameter set as in Fig.~\ref{fig:traj_harmonic}, i.e.~with $x_0 = -1$ and $v(0) = 0$. 
In Fig.~\ref{fig:v_xmax_tau} we present $x_\mathrm{max}$ as a function of the memory time $\tau$. 
Note that for $\tau \to 0$ the maximal position $x_\mathrm{max} = 0$ indicating that the particle stops at the potential minimum $x = 0$. It is due to the fact that in the analyzed strong damping regime $\gamma^2 > 4mk$ all roots $\{\lambda^0_i\}$ of the corresponding characteristic equation are real (see Eq.~\ref{eq:root_memless}) and there are no oscillations but only monotonic decay. Similarly, for the long memory time $\tau \to \infty$ the maximal position $x_\mathrm{max} = 0$. This result follows from the observation that although in such a limit the roots $\lambda^\infty_{2,3}$ are complex (see Eq.~\ref{eq:root_longmem}), the oscillations are dominated by the exponential decay, namely $\lambda_1^\infty > \mathfrak{R}\{\lambda_2^\infty\} = \mathfrak{R}\{\lambda_3^\infty\}$. Finally, the oscillations prevail over the monotonic behavior in the interval $\tau \in (0.09,\ 3.5)$ for which the maximal position $x_\mathrm{max}>0$, so the non-Markovian particle can escape from the hypothetical potential well provided that $x_\mathrm{max} > x_b$, where $x_b$ corresponds to the position of the maximum of the potential barrier.


Let us now consider the particle in a tilted washboard potential $U_p(x) = \sin(2\pi x) - fx$ with $f = 5.65$. 
In the memoryless limit $\tau \to 0$ the considered parameter regime  corresponds to the area in the Risken bistability diagram \cite{Risken1996} where only the locked solutions exist in the phase space and there is no directed transport $\langle v \rangle = 0$. However, when $\tau$ increases, the particle can reach  $x_\mathrm{max} > x_b$ and overcome the potential barrier thanks to the energy recuperated from the bath. Consequently the running trajectories may emerge.
\begin{figure}[t]
    \centering
    \includegraphics{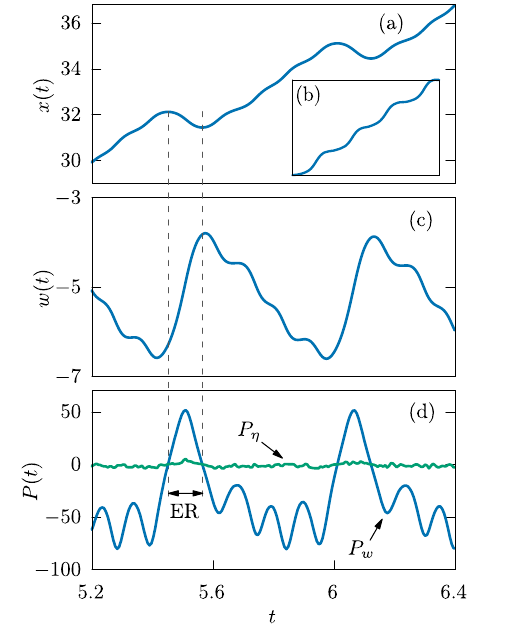}
    \caption{(a): Running trajectory $x(t)$ of the non-Markovian particle dwelling in $U_p(x)$ for $\tau = 0.5$ and $f = 5.65$. (b):~Running trajectory in the memoryless system $\tau = 0$ and $f = 6$. (c):~The friction force $w(t)$. (d): The energy exchange rates $P_w(t)$ and $P_\eta(t)$. ER stands for the energy recuperation. Other parameters are the same as in Fig.~\ref{fig:traj_harmonic} except now $\theta = 0.01$.}
    \label{fig:trajectory}
\end{figure}

A representative example of a single running trajectory of the particle in the tilted periodic potential $U_p(x)$ is presented in Fig.~\ref{fig:trajectory}(a). The particle systematically moves in the positive direction, but every few potential periods it oscillates in the potential well and continues moving forward. This is in contrast to the running trajectories in the memoryless case, where the particle velocity is constantly positive and the regular swings do not occur, see Fig.~\ref{fig:trajectory}(b). To explain the role of the particle oscillatory motion, in Fig.~\ref{fig:trajectory}(c) we plot the time evolution of the friction force $w(t)$ experienced by the particle and simultaneously in Fig.~\ref{fig:trajectory}(d) we present the rate $P_w(t)$ of energy exchange between the system and the surrounding bath. Note that for non-zero temperature $\theta > 0$ the energy exchange is also mediated by thermal noise $\eta(t)$, but as we show in Fig.~\ref{fig:trajectory}(d), the magnitude of $P_\eta(t) = \eta(t)v(t)$ is negligible in comparison to $P_w(t)$.

The particle sliding down the potential $U_p(x)$ slope is initially weakly damped due to the delayed response $w(t)$ of the medium. Consequently, it moves forward and overcomes a few potential barriers. During this motion the particle transfers energy to the bath and $P_w(t) < 0$. When $w(t)$ finally builds up, the particle cannot cross the potential barrier anymore and starts moving in the opposite direction. The friction force, however, again needs time of order $\tau$ to adjust to this change, hence $P_w(t) > 0$, and the particle recuperates energy from the bath. During the backward motion the friction force $w(t)$ lowers, so the particle becomes weakly damped again. This, along with the energy recuperation, allows the particle to finally cross the potential barrier and the whole process restarts. The repeated swings in the potential wells that distinguish the running solutions in the non-Markovian system from those in purely viscous environments are thus necessary for the energy recovery, and consequently for the emergence of directed transport in the studied regime. To sum it up, the delayed friction $w(t)$ allows the particle to recuperate energy stored in the bath and overcome the barriers.

\begin{figure}[t]
	\includegraphics{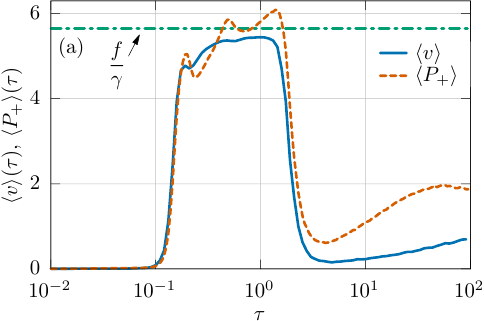}\\
	\includegraphics{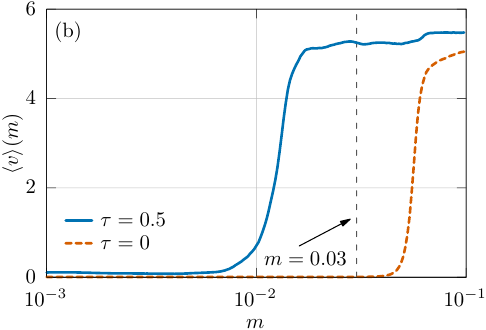}
	\caption{(a): The average velocity $\langle v \rangle$ of the particle in the periodic potential $U_p(x)$ and its mean energy recuperation rate $\langle P_+ \rangle$ presented as a function of the memory time $\tau$. (b): the average velocity $\langle v \rangle$ of the particle versus its mass $m$ in the non-Markovian $\tau = 0.5$ and Markovian $\tau = 0$ setting. Other parameters are the same as in Fig. \ref{fig:trajectory}.}
	\label{fig:v_Pr_tau}
\end{figure}
The latter observation is confirmed in Fig.~\ref{fig:v_Pr_tau}(a) where we depict the average velocity $\langle v \rangle$ of the particle versus the memory time $\tau$, alongside the corresponding mean energy recuperation rate $\langle P_+ \rangle$ given by 
\begin{equation} \label{eq:Pp}
    \langle P_+ \rangle = \lim\limits_{t\to\infty} \langle P_w(t)H[P_w(t)] \rangle,
\end{equation}
where $H(\cdot)$ is the Heaviside step function. We note that in the Markovian limit $\tau \to 0$ there is no directed transport and $\langle v \rangle = 0$. From the inspection of the presented panel it is evident that this phenomenon is memory-induced as for growing $\tau$ the average velocity $\langle v \rangle$ becomes greater than zero. There is also a perfect agreement between the curves $\langle v \rangle(\tau)$ and $\langle P_+ \rangle(\tau)$ so it is clear that the recuperation of energy from the surrounding bath is the mechanism standing behind the emergence of the directed transport in the studied regime. Moreover, we stress that the magnitude of the particle average velocity $\langle v \rangle$ is almost equal to the top speed $f/\gamma$ corresponding to the system with no energy barriers (the free particle limit) which it may achieve in the investigated setup. This proves that the discovered \emph{modus operandi} is not a nuance but rather an efficient mechanism of the directed transport generation. In panel (b) of the same figure we present the characteristics $\langle v \rangle(m)$ to demonstrate that the average velocity  vanishes in the overdamped limit $m \to 0$. It means that the new method of energy recovery related to the inertia of a non-Markovian system lies at the very heart of the mechanism of memory-induced directed transport.

\paragraph{In conclusion,}
in this work we demonstrated that recuperation of energy from the surrounding bath is a generic feature of non-Markovian systems, both in and out of equilibrium (see also End Matter). In doing so we revealed that inertial dynamics alone, in particular in the strong damping regime, can lead to this phenomenon even in the absence of any external forcing. This fact must be contrasted with the previously studied overdamped limit in which energy recuperation requires perturbation of the system. Our findings suggest that energy recovery can be ubiquitous in nature and it may be the \emph{modus operandi} of various phenomena occurring in setups exhibiting memory. As an example we demonstrate that the energy recovery from the surrounding correlated bath can be the mechanism of memory-induced ultrafast directed transport emerging in non-Markovian periodic systems in the strong damping regime. These results vastly improve our understanding of the role of energy exchange between the system and its correlated environment in processes occurring at the microscale, such as dynamics of colloids in viscoelastic fluids or intracellular transport. Moreover, our findings are expected to naturally extend into the realm of open quantum systems since in reality they are inherently non-Markovian. This follows from the fact that the energy of even the simplest quantum system coupled to thermal vacuum diverges when the Markovian (memoryless) limit is imposed \cite{Spiechowicz2021}. 


This work has been supported by National Science Centre (NCN) Grant No. 2024/54/E/ST3/00257 (J.S.).

The data supporting findings of this study are available from the corresponding author upon reasonable request.

\bibliographystyle{apsrev4-2}
\bibliography{bibliography}

\appendix
\section{End Matter \label{endmatter}}

\paragraph{Dimensionless GLE}
The Generalized Langevin Equation describing our final setup in the dimensional form reads
\begin{equation}
	\label{eq:GLE}
    M\dot{v}(t) = -\Gamma\int_0^t K(t-t')v(t')\mathrm{d}t' - U'(x) + \eta(t)
\end{equation}
with
\begin{gather}
    \langle \eta(t) \eta(t') \rangle = \Gamma k_B T K(t-t'),\\
    U(x) = \Delta U \sin(2\pi x/L) - Fx,
\end{gather}
where $k_B$ is the Boltzmann constant, and $T$ is the temperature. Eq.~(\ref{eq:GLE}) can be recast into a dimensionless form by defining the dimensionless position $\hat{x}$ and time $\hat{t}$. One of the possible choices is
\begin{equation}
	\hat{x} = \frac{x}{L}, \quad \hat{t} = \frac{t}{\tau_0}, \quad \tau_0 = \frac{\Gamma L^2}{\Delta U}
\end{equation}
so that the length unit is the spatial period of the potential $U(x)$ and the time unit is related to the timescale of the dynamics of an overdamped particle in the periodic potential. 
Under such scaling the friction coefficient $\Gamma$, barrier of the potential $\Delta U$ and its period $L$ become unity. Other parameters are then replaced by their dimensionless counterparts \cite{Wisniewski2022}
\begin{equation}
    m = \frac{M}{\Gamma \tau_0},\ \hat{\tau} = \frac{\tau}{\tau_0},\ \theta = \frac{k_B T}{\Delta U},\ f = \frac{LF}{\Delta U}.
\end{equation}
For clarity, in the article we omit hats over the rescaled variables.

\paragraph{Derivation of Eq.~(\ref{eq:char_eq})}
The equation describing the dynamics of the friction force $w(t)$ with the memory kernel given by Eq.~(\ref{eq:K}) can be derived via the Markovian embedding procedure and reads \cite{Luczka2005}
\begin{equation} \label{eq:dotw}
    \dot{w}(t) = -\frac{1}{\tau}w(t) - \frac{\gamma}{\tau}\dot{x}(t) - \gamma\varepsilon \ddot{x}(t).
\end{equation}
From Eq.~(\ref{eq:gle}) with $\theta=0$ we also have
\begin{equation} \label{eq:w}
    w(t) = m\ddot{x}(t) + U'(x).
\end{equation}
Differentiating Eq.~(\ref{eq:w}) and comparing it with Eq.~(\ref{eq:dotw}) gives
\begin{equation}
    m\tau\dddot{x} + (\gamma\varepsilon\tau+m)\ddot{x} + \left[U''(x)\tau + \gamma\right]\dot{x} + U'(x) = 0.
\end{equation}
For the harmonic potential $U_h(x)=\frac{1}{2}kx^2$ this gives Eq.~(\ref{eq:char_eq}).
\begin{figure}[t]
    \centering
    \includegraphics{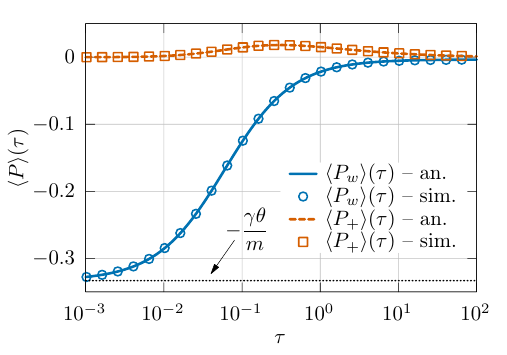}
    \vspace{-1em}
    \caption{The mean rate of energy exchange via the friction force $w(t)$ between the free Brownian particle and its environment for $m=0.03$, $\theta=0.01$ and $\varepsilon=0.01$. Both the net energy exchange rate $\langle P_w \rangle$ and the energy recuperation rate $\langle P_+ \rangle$ are calculated analytically and the results are confirmed by the data from numerical simulations.}
    \label{fig:P_tau}
\end{figure}

\paragraph{Energy recuperation by a free particle in equilibrium}
To prove that energy recuperation is a generic feature of non-Markovian systems, we investigate the exchange of energy between a free Brownian particle and its environment. For the memory kernel given by Eq.~(\ref{eq:K}) and $U[x(t)]\equiv 0$, the equation of motion (\ref{eq:gle}) can be transformed via the Markovian embedding to the following set of equations
\begin{subequations}
\begin{align}
    m\dot{v}(t) &= -\gamma\varepsilon v(t) + z(t) + \eta(t) +\sqrt{2\gamma\varepsilon \theta}\xi_1(t),\\
    \tau\dot{z}(t) &= -\gamma(1-\varepsilon) v(t) -z(t),\\
    \tau\dot{\eta}(t) &= -\eta(t) + \sqrt{2\gamma(1-\varepsilon)\theta}\xi_2(t),
\end{align}
\end{subequations}
where $z(t) = w(t) + \gamma\varepsilon v(t)$ is the delayed part of the friction force $w(t)$, and $\xi_{1,2}(t)$ are independent white noise sources of zero mean obeying $\langle \xi_\alpha(t) \xi_\beta(t') \rangle = \delta_{\alpha\beta}\delta(t-t')$ with $\alpha,\beta\in\{1,2\}$. The corresponding Fokker-Planck equation for the probability density $p(v, z, \eta, t)$ reads \cite{Risken1996}
\begin{widetext}
\begin{equation} \label{eq:FP}
    \frac{\partial p}{\partial t} = \frac{\partial}{\partial v}\left[\left(\frac{\gamma\varepsilon}{m}v - \frac{1}{m}z - \frac{1}{m}\eta\right)p\right] + \frac{\partial}{\partial z}\left[\left(\frac{\gamma(1-\varepsilon)}{\tau} v +\frac{1}{\tau} z\right)p\right] + \frac{\partial}{\partial \eta}\left[\frac{1}{\tau}\eta p\right] + \frac{\gamma\varepsilon \theta}{m^2}\frac{\partial^2 p}{\partial v^2} + \frac{\gamma(1-\varepsilon)\theta}{\tau^2} \frac{\partial^2 p}{\partial \eta^2}.
\end{equation}
\end{widetext}
The stationary solution (for $\frac{\partial p}{\partial t} = 0$) of Eq.~(\ref{eq:FP}) is a Gaussian distribution
\begin{equation}
    p(v, z, \eta) = \frac{1}{\sqrt{(2\pi)^3\det[S]}}\exp\left[-\frac{1}{2}\mathbf{X}^T S^{-1}\mathbf{X}\right],
\end{equation}
where $\mathbf{X}^T = \begin{bmatrix}v & z & \eta \end{bmatrix}$ and $S$ is the covariance matrix reading
\begin{equation}
    S = \!\!\setlength{\arraycolsep}{1.6pt}
    \begin{bmatrix}
        \langle v^2 \rangle & \langle v z \rangle & \langle v\eta \rangle \\
        \langle v z \rangle & \langle z^2 \rangle & \langle z\eta \rangle \\
        \langle v\eta \rangle & \langle z\eta \rangle & \langle\eta^2\rangle
    \end{bmatrix} \!\!
    =\! \frac{\theta}{m}\frac{\tilde{\gamma}}{1+r}\!\begin{bmatrix}
        \frac{1+r}{\tilde{\gamma}} & -1 & 1 \\
        -1 & \tilde{\gamma} & -\frac{\tilde{\gamma}}{2} \\
        1 & -\frac{\tilde{\gamma}}{2} & \frac{m(1+r)}{\tau}
    \end{bmatrix}\!\!,
\end{equation}
where $\tilde{\gamma} = \gamma(1-\varepsilon)$ and $r = \frac{\gamma\tau(1+\varepsilon)}{2m}$.
The mean rate of energy exchange via the friction force $w(t)$ can then be retrieved from $S$ and reads
\begin{equation} \label{eq:meanPw}
    \langle P_w \rangle = \langle wv\rangle = \langle vz \rangle - \gamma\varepsilon\langle v^2 \rangle = -\frac{\gamma \theta}{m}\frac{1+\varepsilon r}{1+r}.
\end{equation}
The rate of energy recuperation can be calculated by integrating $p(v, z, \eta)$ over the region where $wv > 0$, i.e.
\begin{align} \label{eq:meanPp}
    \langle P_+ \rangle &= 2\int_{0}^{\infty}\! \mathrm{d}v \int_{\gamma\varepsilon v}^\infty\! \mathrm{d}z \int_{-\infty}^\infty\! \mathrm{d}\eta \ (z-\gamma\varepsilon v)v p(v, z, \eta) = \notag \\
    &= -\langle P_w \rangle \frac{\frac{(1-\varepsilon)\sqrt{r}}{1+\varepsilon r}-\arctan\left(\frac{(1-\varepsilon)\sqrt{r}}{1+\varepsilon r}\right)}{\pi}.
\end{align}
The factor 2 in the integral follows from the fact that $wv > 0$ for both $\{v>0,\ z > \gamma\varepsilon v\}$ and $\{v < 0,\ z < \gamma\varepsilon v\}$, and from the symmetry of $p(v, z, \eta)$. $\langle P_+\rangle$ vanishes both in the memoryless case $\tau = 0$ and in the long memory limit $\tau\to \infty$, but it is positive otherwise, since $\frac{(1-\varepsilon)\sqrt{r}}{1+\varepsilon r} > \arctan\left(\frac{(1-\varepsilon)\sqrt{r}}{1+\varepsilon r}\right)$ for $\tau > 0$. This means that the free particle recuperates energy from the environment whenever $\tau > 0$.

In Fig.~\ref{fig:P_tau} we illustrate the rate of energy exchange as a function of the memory time $\tau$. In accordance with Eq.~(\ref{eq:meanPw}), the magnitude of net energy exchange rate is lower than the one in the memoryless limit ${\langle P_w \rangle(\tau=0)} = -\gamma \langle v^2\rangle = -\gamma \theta/m$, but it remains negative since the system is dissipative. The product $w(t)v(t)$, however, can be temporally positive due to the delayed response of the friction force to the particle dynamics, resulting in $\langle P_+ \rangle > 0$. The analytical predictions perfectly match the results of numerical simulations.

Finally, we note that for the overdamped limit $m \to 0$ the mean energy exchange rate $\langle P_w \rangle$ diverges to infinity. This follows from the fact that in such a case the velocity fluctuations as well as the kinetic energy are not well defined. $\langle P_w \rangle$ does not exist individually, but only in combination with $\langle P_\eta \rangle$ \cite{Sekimoto2010, Jung2011}. Therefore to calculate the energy recuperation rate in non-Markovian systems one has to consider an energetically viable model taking into account inertia of the setup, i.e. analyze the strong damping instead of the overdamped regime.

\end{document}